\documentclass[preprint,review,12pt]{elsarticle}
\usepackage{graphicx}
\usepackage{amssymb}
\usepackage{amsmath}
\usepackage{bm}
\usepackage[version=4]{mhchem}
\journal{Journal Name}

\begin{document}

\begin{frontmatter}

\title{The Frogs Method: A Bio-inspired Algorithm For Micromagnetic Dynamics Skyrmion Data Extraction}

\author{J. P. A. Mendon\c ca}
\author{I. A. Santece}
\author{M. G. Monteiro Jr.}
\author{S. A. Leonel}
\author{P. Z. Coura}
\author{F. Sato}

\address{Departamento de F\' isica, Universidade Federal de Juiz de Fora (UFJF), Brazil}

\begin{abstract}
In this work we present a new bio-inspired method (using an adapted, frog based Particle Swarm Optimization) that can track a skyrmion's position in micromagnetic simulation. The structure and position of skyrmions in a given system is usually determined either by visual impression through a vector plot, or by averaging over magnetization and topological charge. Due to the inherent difficulty of acquiring this information in discrete systems, and also in systems with finite boundaries, we present a method which is able to reliably track a wide array of different topological structures, while also providing additional data about internal degrees of freedom, in both stationary and dynamic cases.
\end{abstract}


\end{frontmatter}

\section{Introduction}
\label{introduction}
Skyrmions first appeared in particle physics \cite{Skyrme1962}, and are topological excitations of a system that behave as quasiparticles. From recent experimental results, we know that many systems in condensed matter \cite{wright1989crystalline,ho1998spinor,choi2015corrigendum, Machida} and materials science \cite{morice2018pseudo,nagaosa2013topological,sampaio2013nucleation} can present skyrmions. In particular, they appear as stable states in some magnetic nanostructures, where they are studied in the context of fundamental nanomagnetism and also in employing skyrmion quasiparticles in spintronic devices \cite{fert2017magnetic,doi:10.1142/9789814287005_0015}. Regarding the latter, skyrmions represent a great alternative for the future of data storage and for logic processing devices \cite{fert_cros_sampaio_2013,Kiselev_2011}.

Experimentally, skyrmions have first been observed for systems under the effect of external magnetic fields \cite{Muhlbauer915,Romming636}. With the progress of experimental techniques, skyrmions have been stabilized at room temperature even without external magnetic fields \cite{BAZEIA20161947,zhangsenfu}. Usually, skyrmions can appear through two known mechanisms in magnetic systems: by the Chiral Magnetic Effect (CME) of a material, which may result in the creation of a lattice of stable and localized solitons \cite{nagaosa2013topological,PhysRevB.Dai}, and by the presence of an induced Dzyaloshinskii-Moriya (DM) coupling \cite{DZYALOSHINSKY1958241,MORIYA} corresponding to a strongly antisymmetrical exchange interaction among magnetic moments in the interface of two materials \cite{CREPIEUX1998341,PhysRevB.interface,PhysRevLett.Bogdanov}. In both cases, the fundamental property that rules the appearance of stable skyrmions is a breaking of inversion symmetry in systems with very strong spin-orbit coupling \cite{Kanazawa_2016}.

Because skyrmions are topological entities, they can be handled in a theoretical manner by continuous models \cite{büttner_lemesh_beach_2018,lee_moon_rim_2001}. From the computational point of view, the mathematical treatment is done in a discretized and numerical way, typically being addressed by the micromagnetic approach\cite{brown1963micromagnetics}. Any comparison between theoretical, experimental and simulation data is strongly related to our capacity of obtaining properly comparable measurements from the three approaches. The quality and credibility of these measurements is the core to our confidence in the results. Since more general software packages do not aim to address every kind of specific problem, it is not always easy to extract data from simulations at the same time as they are running. Even for some straightforward and fundamental measurements such as the skyrmion positions, there is no consensus in literature as to what is the most appropriate technique (given that it can be done in many different ways \cite{PAPANICOLAOU1991425,KOMINEAS199681,Moutafis}), and also what is the error associated to each numerical approach.

During the last years, bio-inspired algorithms are appearing in many areas as a good
alternative to more traditional methods, giving rapid and satisfying solutions to very complex computational problems \cite{olariu_zomaya_2006,mukhopadhyay_2014,gong2015bio}. Heuristics based in biological systems have already been used in other fundamental sciences, being able to predict optimal paths \cite{gong2015bio}, geometric configurations of complex molecules \cite{heberle_azevedo_2011,LI201470}, and even to model potentials of interactions \cite{li_kermode_vita_2015,chmiela_2017}.   

Observing the difficulties of traditional methods in describing the position and internal degrees of freedom of skyrmions, we develop a rather flexible method that can describe many topological variants including skyrmions. To this purpose, in this work we present a technique to extract additional data from traditional micromagnetic simulations using a frogs inspired method. We focus our attention in obtaining skyrmion trajectories and other internal variables integrated over time during simulations, generating more coherent data sets and giving access into previously inaccessible values over the course of the magnetization dynamics. In the Methodology (section \ref{methodology}), we discuss the initial assumptions used in our micromagnetic calculations and the theory behind the Frogs Method. In the Results (section \ref{results}), we present the obtained data for some test models, highlighting the advantages of the methodology presented, when compared to previously reported techniques.

\section{Methodology}
\label{methodology}
In the micromagnetic model \cite{brown1963micromagnetics}, a given atomic system is redefined as a cluster of cells, each containing N atomic magnetic moments ($\bm\mu$) whose  directions are roughly the same (i.e the angle $\theta$ among every $\bm\mu$ is very small and varies smoothly across cells). Considering these assumptions, the material can then be described by the average density of atomic moments contained in each cell. This average, for appropriate system size, is then approximately a continuous function of both time and cell position, $\textbf{M}(\textbf{r},t)$.
Micromagnetism allows us to deal with systems of nano- and mesoscopic size, by reducing the number of effective interactions that would otherwise have to be calculated in a fully quantum mechanical description. With this, there is a considerate gain in simulation time and a decrease of the use of system RAM memory. 

Since in practice the structure is composed of a cluster of cells, it is very convenient to employ a finite difference scheme to spatially decompose the field $\bm{M}$ into a set of local magnetization vectors, located in each cell center, given by:

\begin{equation}
    \label{eq:1}
    \textbf{M}_{i}=\frac{1}{V_{cel}} \sum_k \bm{\mu}_{k}= \frac{\textbf{m}_i}{V_{cel}}
\end{equation}
 where $V_{cel}$ is the cell volume and $\bm{m}_i$ the resulting magnetic moment in cell $i$.

The cell partition size must be bound within the exchange length  $\lambda_{ex}=\sqrt{\frac{2A}{\mu_0 M_{s}^2}}$, which represents the spatial scale at which the (local) exchange interactions becomes negligible, therefore degrading the resolution of our model. We choose a cubic cell of size $a_0$ such that $a_0 \leq \lambda_{ex}$ to ensure that the atomic moments contained in each cell will be approximately aligned i.e, the short range exchange interactions dominates magnetostatic effects. $A$ represents the exchange stiffness, $\mu_0$ is the vacuum permeability and $M_s$ is the saturation magnetization.

 We also assume the cells have the same number of atomic moments in average, therefore having a structure where the magnetization vectors have a constant modulus, varying only their directions. The latter approximation only fails in continuous boundaries that cannot be properly build up from a cubic lattice. Nevertheless, a smaller lattice parameter $a$ might still be able to reasonably describe such systems.

This approach is very common in studying topological excitations of magnetic systems, like in the case of skyrmions \cite{TOSCANO2019171,Boulle2016,Zhou2015,Bhattacharya2016} or vortices \cite{MOREIRA2017252,PetitWatelot2012,Brando2014}. As mentioned, this is due to the typical size range of structures lying in the interval from nanometers to several micrometers. In this size range, an atomistic approach would require handling as many as $(10^{10})^2$ interactions in average (e.g, for the widely used Cobalt, we have a numerical density of $\approx 9.1 \times 10^{10} \mu m^{-3}$). The necessity of simulating dozens or even hundreds of different initial conditions and constraints at a time for each such system puts the quantum mechanical approach far beyond the capabilities of available hardware.

As skyrmions are stable and topologically protected, it's common in literature to treat them by the continuous theoretical approach presented by Thiele \cite{thiele} for magnetic bubbles. In this case, the dynamics of the magnetization field, given by the L.L.G. equation \cite{LLG}, is reduced by introducing the transformation $\textbf{m}(\textbf{r},t)=\textbf{m}_0(\textbf{r}-\textbf{R}(t))$, where $\textbf{R}(t)$ is the skyrmion position. This is equivalent to a Galilei transformation, where the quasiparticle becomes the origin of the system of coordinates, and the evolution of the field $\textbf{m}$ can be described by a simple translation of $\textbf{m}_0(\textbf{r})=\textbf{m}(\textbf{r},0)$ over the trajectory $\textbf{R}(t)$.

Following the method proposed in  \cite{BAZEIA20161947}, 
we use the ansatz of a skyrmion in the form $\textbf{m}_0=\sin(\frac{\pi}{2}\phi+\delta)\hat{z}+\cos(\frac{\pi}{2}\phi+\delta)\hat{\theta}$, where we use a cylindrical coordinate system, i.e. $\textbf{r}=r\hat{r}(\theta)+z\hat{z}$, $\delta$ is a phase angle and $\phi$ is given by

\begin{equation}
    \phi(r) = \frac{1-\left(\frac{r}{R_S}\right)^{2/(1-s)}}
    {1+\left(\frac{r}{R_S}\right)^{2/(1-s)}}
    \label{eq.skyr}
\end{equation}    
   
In this ansatz, $s\in[0;1)$ is a dimensionless parameter that can be associated with the strength of the magnetic interactions in the material and $R_S \in(0;\infty]$ is the skyrmion radius. From now on we will consider a null phase $\delta=0$, but it can also be included in the optimization process for more complex skyrmion configurations. 

In this setup, the full dynamics of a given magnetic system is completely known if we can determine $\textbf{R}(t)=x_S(t)\hat{x}+y_S(t)\hat{y}$, $R_s$ and $s$. In more general approaches, it could happen that boundary conditions, external driving forces and curvature change the skyrmion shape, and in this case we have both $R_s$ and $s$ as functions of time, or an even more complex ansatz would be needed to properly describe the skyrmion shape.

A typical micromagnetic simulation (such as the ones performed by popular software packages OOMMF \cite{OOMMF} and MuMax$^3$ \cite{MuMax}) usually gives as output the dataset $\textbf{m}_i(t_n)$ for each moment $i$ in given time steps $t_n$. We can then focus our efforts in extracting from this kind of data sets the values of $\textbf{R}$, $R_S$ and $s$. For that, we introduce a bio-inspired optimization algorithm that has its meta-heuristic based in a frog population searching for food abundant regions. In this methodology, closely related to the classical Particle Swarm Optimization (PSO) \cite{clerc2012standard,bratton2007defining}, we explore the following assumptions over our motivating system, i.e, a population of frogs:

  \begin{enumerate}
            \item The frogs are able to change their positions all at once, leaping to randomized places from each one's vicinity, in search for a better place to hunt food.
            \item Each frog uses the information given by the sounds emitted by others (\textit{croaks}). Through this means of communication, in analogy to the PSO, a single frog has by itself the information of the best position already known by the whole group (\textit{army}) and uses it to leap onto a better direction.
            \item When jumping, a frog weights its own individual, fully randomized leap with one leap that is entirely based on the \textit{army's} information. This is made by taking a mean value from the whole frog set, weighted by an adjustable parameter in the model.
            \item The maximum size of a frog leap is limited to its physical capabilities. With this in mind, an additional adjustable parameter is introduced in order to give each frog a limit of single leap length.
            \item If a leap puts a given frog in a place that is worse than the last one, the frog will automatically undo its own leap. For all effects and purposes, it will be as if the frog didn't leap at the present time step. 
        \end{enumerate}

We highlight that the item 1 of our assumptions introduces a key difference from the traditional PSO, that is, the frogs will have a randomized trajectory. Doing so, the model acts also in analogy to the well known Metropolis algorithm \cite{hastings,metropolis,spall_2003}. In a sense, this means that the Frog Method, as in Markov Chain Monte Carlo optimization, can potentially cover all of configuration space if given enough time. The advantage brought by the frogs in comparison to that class of methods, even against Multi-path Monte Carlo \cite{Raki2016}, is that the multi-agent approach of frogs enable also that the agents exchange information, thereby cutting down on computational time in many cases. In particular, it allows that different local minima are weighted in direct comparison during data analysis.
PSO has been previously tested and its efficiency has been demonstrated in a series of sample cases, as can be seen in \cite{vesterstrom2004comparative} and \cite{zhu2011particle}. Some interesting applications can even be handled by a combination of PSO coupled to Monte Carlo, using MC output as an initial configuration to a PSO refinement, as done in \cite{gong2017monte}. Even more so, combining ideas from PSO and other methods, state of the art algorithms with high adaptability and efficiency are emerging, like in the case of the Elephant Herding Optimization algorithm \cite{elhosseini2019performance}.

Giving these assumptions, we can assign to a given frog $f$ the 4D position vector (in configuration space) $\textbf{x}^f(t)=(x^{1,f}_t,x^{2,f}_t,x^{3,f}_t,x^{4,f}_t)$, with $x^1:=x_S$, $x^2:=y_S$, $x^3:=R_S$ and $x^4:=s$. The time evolution of a given configuration is then modelled by the change in the frogs position vectors, that will be processed just as in PSO, using $x^{i,f}_{t+1}=x^{i,f}_t+v^{i,f}_t$. Note that, in this model, we are treating the discrete optimization ``time'' as dimensionless ($\Delta t=1$), with $v^{i,f}_t$ having the same units as $x^{i,f}_t$. Also, the concept of ``time'' we are dealing with here is related to the optimization process in configuration space, and is not to be confused with the continuum real time that appears on both L.L.G. and Thiele equations.

To determine the optimization leaps given by the frogs, we will use the following formulae, based on our assumptions listed before:
        
        \begin{equation}
            v^{i,f}_{t}=\frac{L^i_{max}}{\rho+1}
            \left(
            \textbf{Rand}[-1,1]+
            \rho \textbf{Rand}[0,1]\left(\frac{g_t^i-x^{i,f}_t}{|g_t^i-x^{i,f}_t|}\right)
            \right).
        \end{equation}
        
$L^i_{max}$ is the max leap length for the $i$-th degree of freedom and $\rho$ indicates how much the frog will trust the \textit{army} opinion in detriment of the random step, both adjustable parameters presented in the assumptions. $\textbf{Rand}[a,b]$ represents a random number generated in the interval $[a,b]$ each time this calculation is done and $g_t^i$ represents the value for the $i$-th degree of freedom of the best position known by the \textit{army} in the instant $t$. 
        
At each step, it will be the case to one frog that $g_t^i=x^{i,s}_t$, which can cause problems with our $v$ formulae as they are defined. Additional caution in calculations has to be taken to evade divisions by 0.

To evaluate the step in configuration space, the frog has to calculate the quality of its new proposed configuration in order to compare with the previous ones. This is the problem of finding a fit, i.e. a function that says how good a solution is. Here, we want to find the values of $(x_S,y_S,R_S,s)$ that best represent the $\textbf{m}_i$ given by a previous micromagnetic simulation. To do so, we propose the fitting function $f(\textbf{x}^f)$ as:

\begin{equation}
    f(x_S,y_S,R_S,s)=\sum_i{\left|\textbf{m}_i \cdot \hat{z} -
    \sin\left(\frac{\pi}{2}\times\frac{1-\left(\frac{(x_i-x_s)^2+(y_i-y_s)^2}{R_S^2}\right)^{\frac{1}{1-s}}}
    {1+\left(\frac{(x_i-x_s)^2+(y_i-y_s)^2}{R_S^2}\right)^{\frac{1}{1-s}}}\right)\right|}.
    \label{eq.fitness}
\end{equation}

Where $(x_i,y_i)$ is the position of the $i$-th micromagnetic cell. Expression (\ref{eq.fitness}) represents the total error in predicting the value of $m_z$, integrated over every cell. With this, the best position will be the one with the lower value of $f$. The full error integrating the difference in the three dimensions of $\textbf{m}$ can also be used, but it will require more computational time, and this choice of $f$ has shown to be good enough in our studied systems (as will be seen in section \ref{results}).

For all the calculations shown in section \ref{results}, we used $10$ frogs that scouted for the skyrmion during $50$ steps. As a maximum leap length, we used $L_{max}^x=L_{max}^y=L_{max}^{R_s}=0.5a_0$, $L_{max}^s=0.5$. $\rho$ has been chosen as 1.

In order to compare the obtained results of this method with others that can be found in literature, we estimated $x_S$ and $y_S$ by other two previously reported methods \cite{Moutafis}. The first one is to determine the skyrmion core position by equating it to the location of the mean $\left<m_z\right>$ of the structure, when $m_z$ is treated in the sense of distributions. It can be done by the following equation:

\begin{equation}
    x_{<m_z>}=\frac{\int x(m_z-1)dV}{\int (m_z-1)dV};\\ \hspace{0.5cm}
    y_{<m_z>}=\frac{\int y(m_z-1)dV}{\int (m_z-1)dV}.
    \label{eq:mz}
\end{equation}

A second way is to first calculate the density of topological charge, or winding number $n$, by the formula:

\begin{equation}
    n=\frac{1}{2} \epsilon_{\mu\nu} (\partial_\nu \textbf{m} \times \partial_\mu \textbf{m})\cdot\textbf{m}
\end{equation}

Where $\mu$ and $\nu$ sum over $x$, $y$ and $z$. After that, we can equate the skyrmion center to the position of the mean of topological charge distribution of the system. This can be done by the following equation:

\begin{equation}
    x_{<n>}=\frac{\int x n dV}{\int n dV};\\ \hspace{0.5cm}
    y_{<n>}=\frac{\int y n dV}{\int n dV}.
    \label{eq:n}
\end{equation}

In both methods, the micromagnetic approximation is used, by changing the integral of $dV$ into a sum over cell elements $i$. In the discrete systems, all the derivatives are treated as centered finite differences and all boundary cells have been unconsidered.

\begin{figure}[h]
    \centering
    \includegraphics[width=7cm]{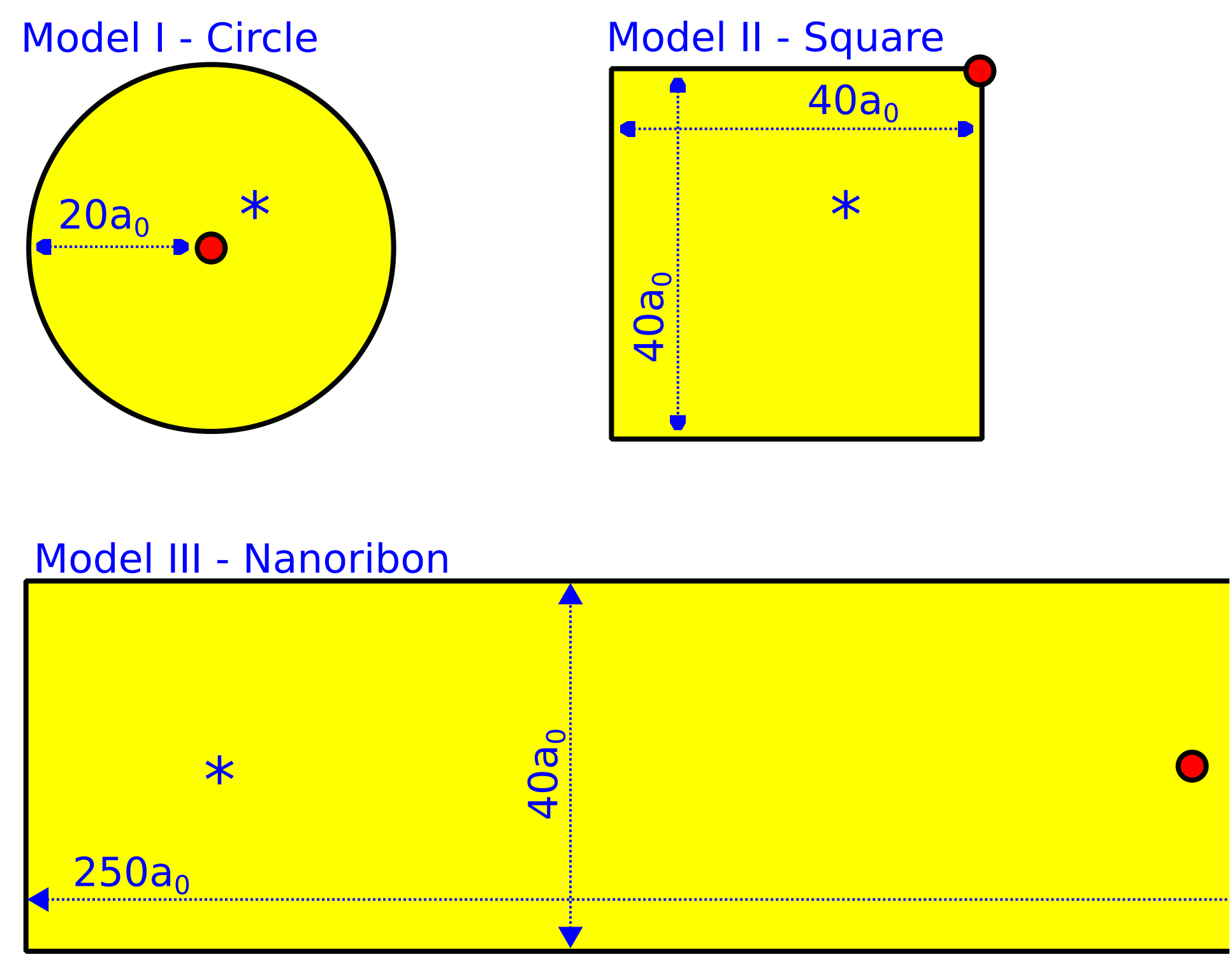}
    \caption{Three models used in our micromagnetic simulations. The blue asterisk represent the initial position of the skyrmion center and the red circle represent the origin of the coordinate system adopted (in the center of the model for I and III, but in the vertex of the square in model II).}
    \label{fig:models}
\end{figure}

To apply the Frogs, the Center of $m_z$ and the Center of $n$ techniques, we generated three model micromagnetic structures, in which we evaluated the magnetization evolution over time starting from a skyrmion configuration. The model geometries and initial skyrmion positions can be seen in Fig. \ref{fig:models}. In model III, a current density was also used as driver for the skyrmion motion. The simulation parameters are set for those of a standard \ce{CoPt} interface with z-axis anisotropy $K$ and an induced Dzyaloshinskii$-$Moriya interaction $D$ : Ms = $5.8 \times 10^{5}$[A/m], $A = 15 \times 10^{-12}$ [J/m], $K = 12.0 \times 10^{5}$ [J/m$^3$], $D = 4.0 \times 10^{-3}$ [J/m$^2$]. During dynamics simulations of the LLG equation, a damping parameter of $\alpha = 0.1$ and cell size parameter of $a = 2.0 \times 10^{-9}$[m] were used, with a timestep of $\Delta t \approx 4.3 \times 10^{-15}$[s]. In the dynamics of model III, a Spin-Transfer Torque driving current \cite{STT} of $1.0 \times 10^{12}$[A/m$^2$] was used, with a non-adiabatic momentum transfer parameter of $\beta = 0.35$ and a mean polarization of $P = 0.7$. 

Our implementation of the Frogs Method has been done using Fortran 90, in serial programming. A full pseudo-code describing the implementation can be found in \ref{pseudocode}. The micromagnetic scheme used in our simulations is due to the authors own group, and has been previously used to describe a wide arrange of magnetic systems \cite{MOREIRA2017252, PAIXAO2018639, TOSCANO2019171}.

\section{Results}
\label{results}

\begin{figure}[h]
    \centering
    \includegraphics[width=7cm]{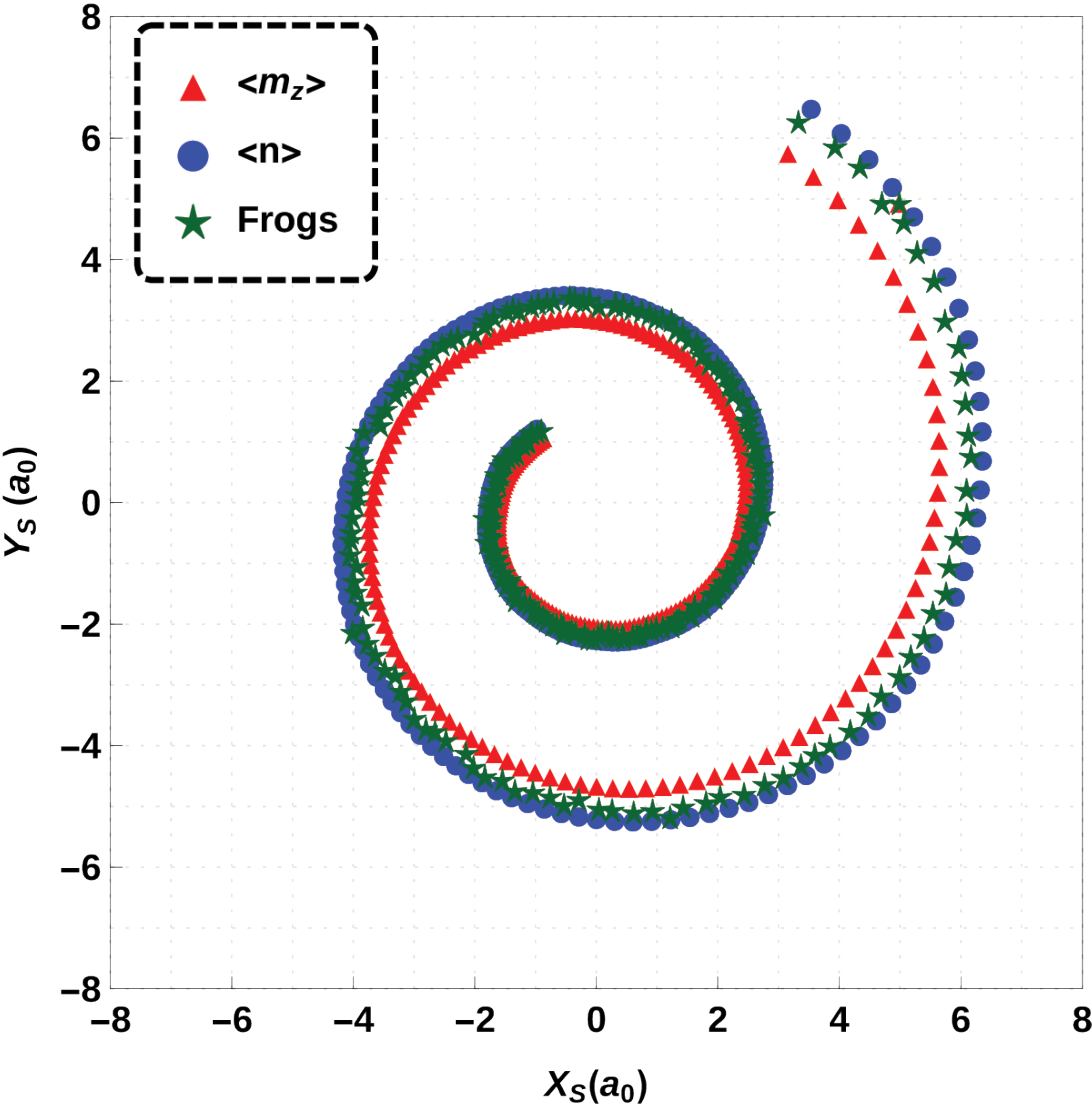}
    \caption{Trajectories obtained for the three tested methodologies in the model I.}
    \label{fig:trj_circulo}
\end{figure}

\begin{figure}[h]
    \centering
    \includegraphics[width=7cm]{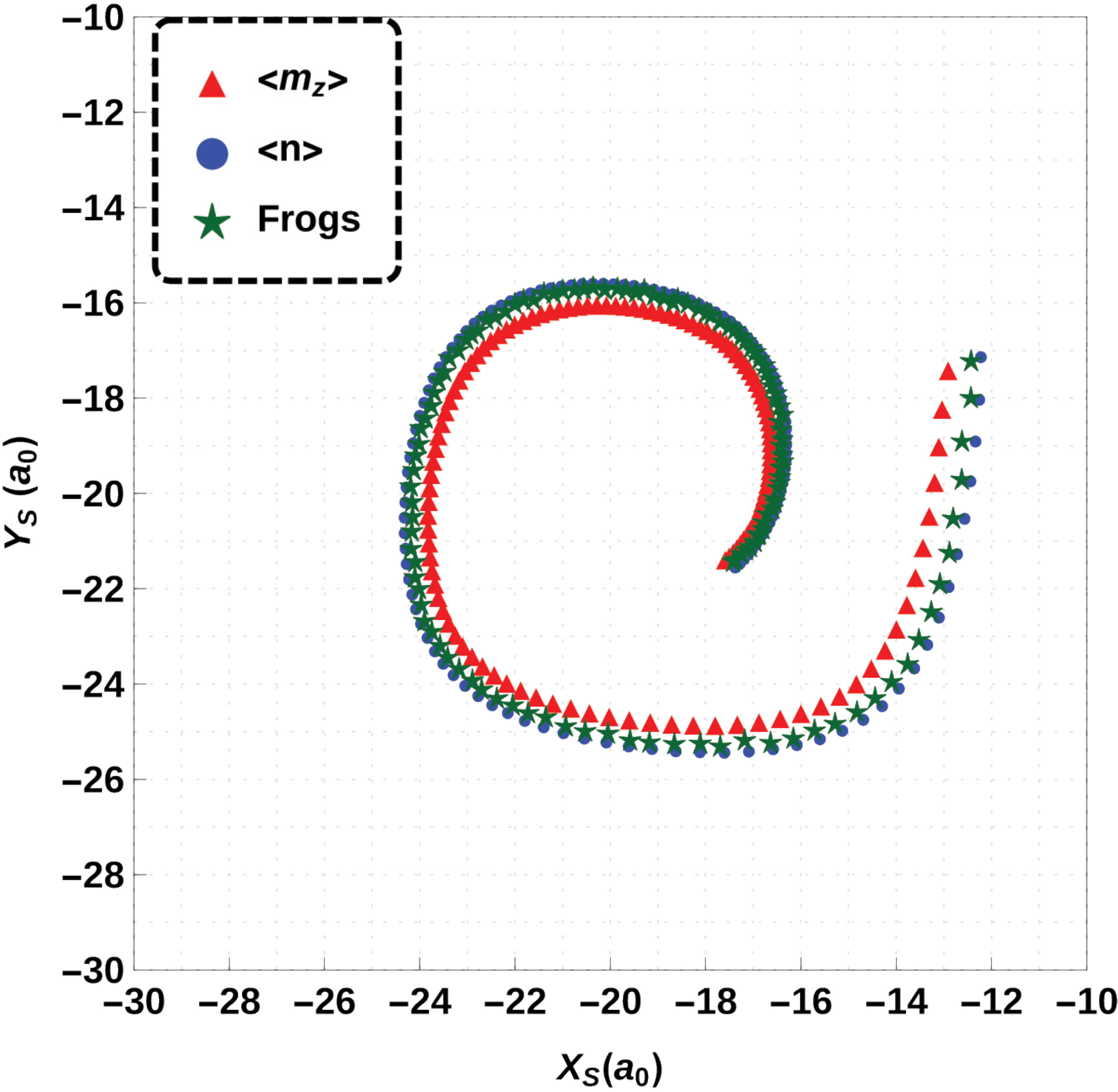}
    \caption{Trajectories obtained for the three tested methodologies in the model II.}
    \label{fig:trj_quadrado}
\end{figure}

\begin{figure}[h]
    \centering
    \includegraphics[width=7cm]{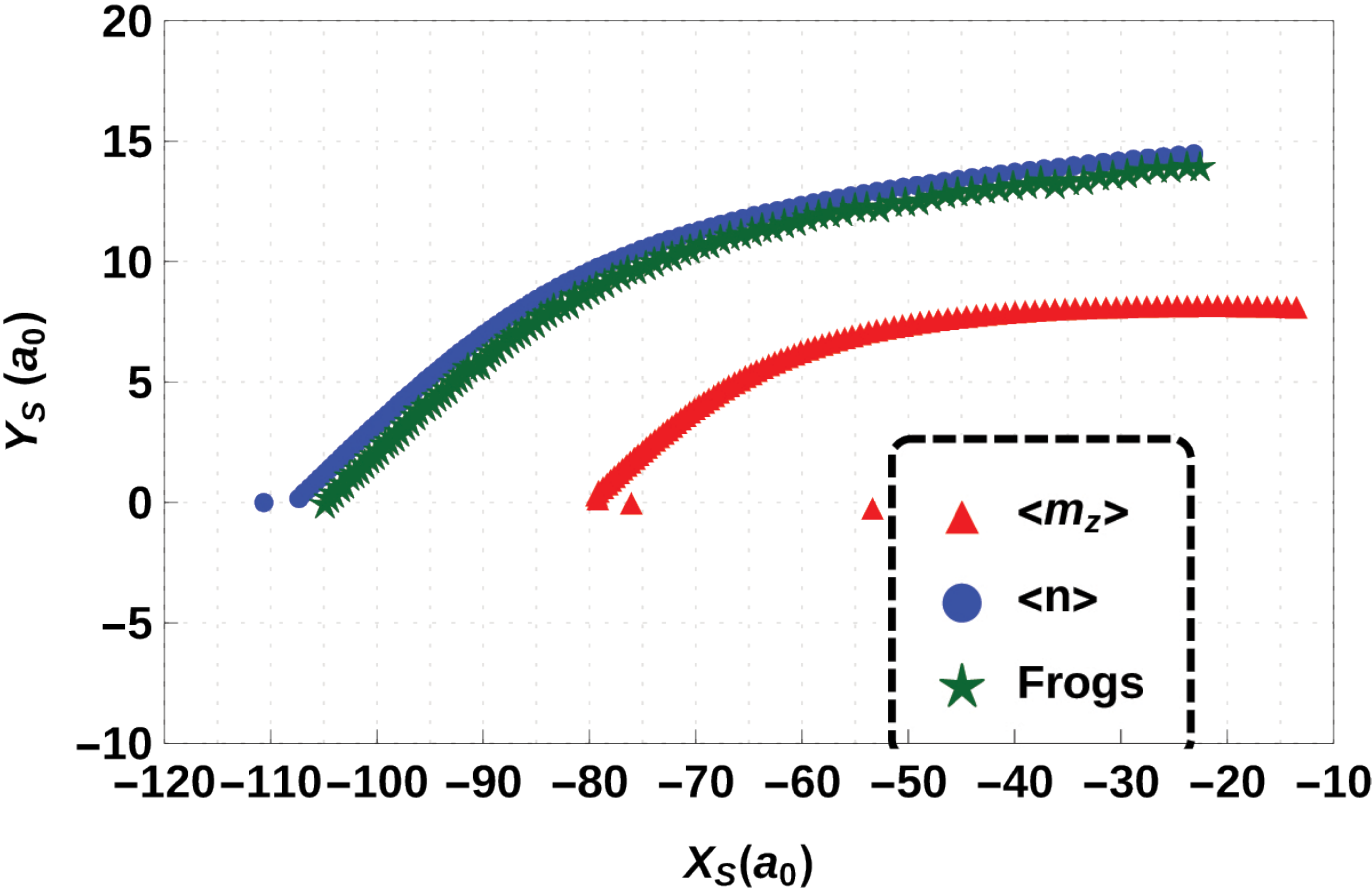}
    \caption{Trajectories obtained for the three tested methodologies in the model III.}
    \label{fig:trj_fita}
\end{figure}

The three proposed models were built as described in Fig. \ref{fig:models}. We proceed with plotting the obtained trajectories for the three models and comparing the obtained results with the Frogs Method and the others. The obtained trajectories can be seen in figure \ref{fig:trj_circulo} for model I, in figure \ref{fig:trj_quadrado} for model II and in figure \ref{fig:trj_fita} for model III.

In models I and II, even with very similar outcomes, we can begin to see the nature of the behavior of each method. The Frogs Method usually returns a position in between the predicted position from the expected values of $m_z$ and $n$, with the one from $m_z$ closer to the center of the structure. Even with the geometric center of model II not coinciding with the origin of the coordinate system, the results of taking the position as the position of mean value $\left<m_z\right>$ seems to present a systemic deviation in the direction of the structure geometric center. 

\begin{figure}[h]
    \centering
    \includegraphics[width=7cm]{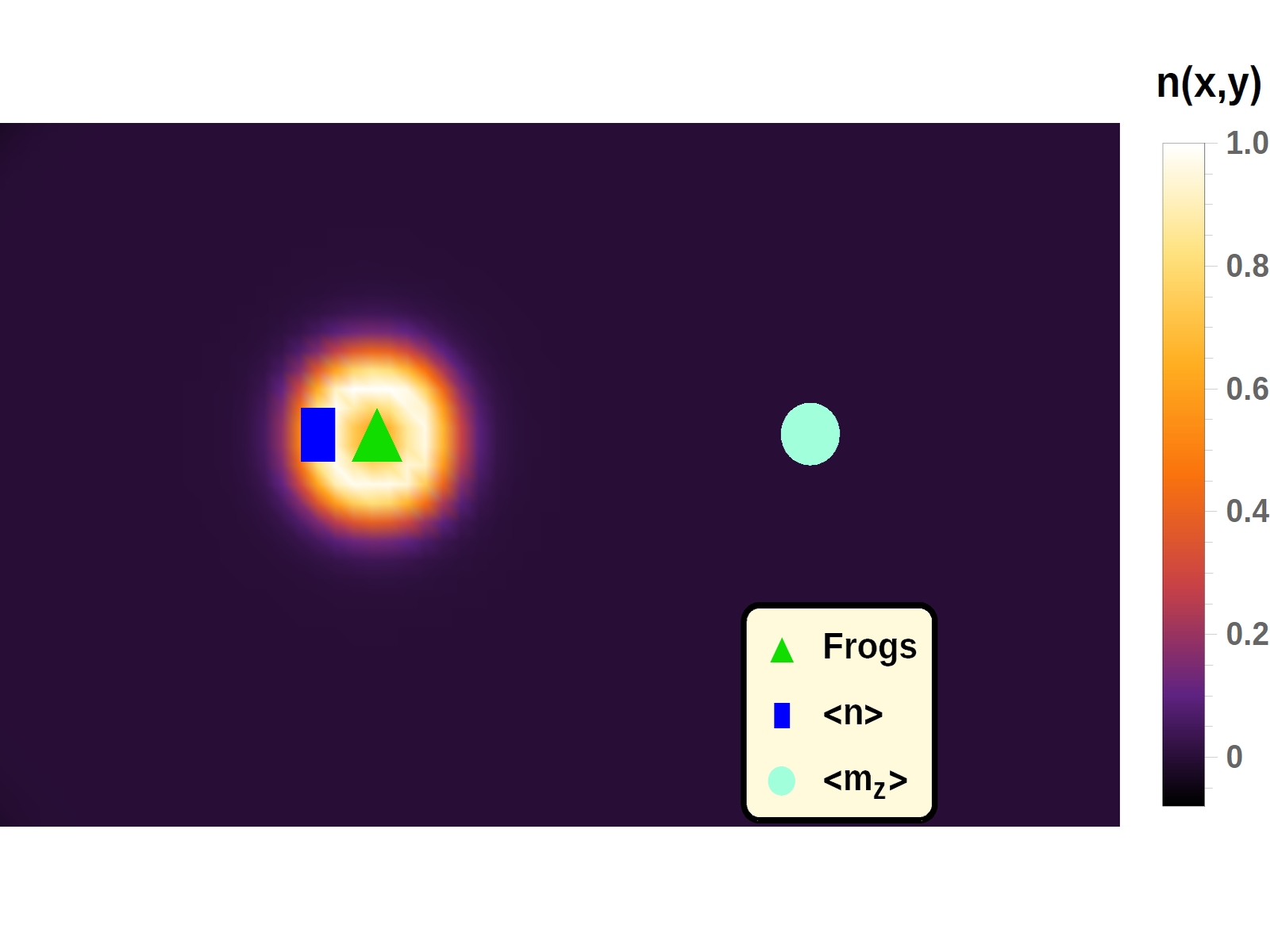}
    \caption{Graphic representations of the predicted position of the skyrmion by the different methods for model III. The background is a density plot of the topological charge $n$ across the ribbon, and markers correspond to the prediction by each method.}
    \label{fig:frame50}
\end{figure}

This divergence becomes notable when analyzing the trajectory for model III, where the initial stable values of $x_S$ using $m_z$ are approximately $25a_0$ higher than the same value from $n$ or from the frogs method. From the frame in figure \ref{fig:frame50}, we can see that in fact the $m_z$ based prediction is far from the visual skyrmion position. Also in this figure, we can see that even the prediction from $n$ deviates slightly from the actual center of the skyrmion. 

From the definition of both skyrmion positions, based on $m_z$ and $n$, we can see that near a boundary, the values will tend either away from, or towards the edges, due to translation symmetry breaking that produce a net topological charge. This effect in turn pulls the position of both mean values out of the actual skyrmion core, coinciding only in the limiting case of a plane.

Even though the Frogs Method introduces noise through random number trials, which is not present in the other methodologies, it does not show any position bias, either towards to, or away from edges. This is because each frog purposely considers the magnetization changes close to boundaries in the integration, repeatedly during each leap, by use of the fitting function.

Regarding precision, the frogs results can be systematically improved by changing the parameters and averaging over many calculations of the same value using different seeds. However, even within a single attempt at predicting the position, each obtained value from the frogs comes with the calculated fitting function defined in eq. \ref{eq.fitness}, therefore giving a direct notion of whether or not, and by how much, can the measurement be trusted. The combination of these factors leads to a straightforward self-consistent optimization algorithm that is likely to be better suited in describing more complex systems, where the assumption that $\bm{m}_i$ and $n$ are smooth distributions during all of dynamics breaks down, and also in the cases where they do not have clearly defined peaks, such as skyrmion lattices or during the interaction of one or more skyrmions or anti-skyrmions.

In addition, our tests pointed out that the typical deviation in the skyrmion tracking position for the same input $\textbf{m}_i$ and with the parameters used in this work, i.e only modifying the random numbers, is approximately $0.1a_0$, which represents a remarkable achievement in precision, compared against results from the other methods. 

With every studied case, the dynamics started with an ideal skyrmion as described by equation \ref{eq.skyr}, using $R_S=4.00a_0$ and $s=0.50$. We additionally used the Frogs Method to directly track time evolution of these parameters, since they are inaccessible to the other methods.

\begin{figure}[h]
    \centering
    \includegraphics[width=7cm]{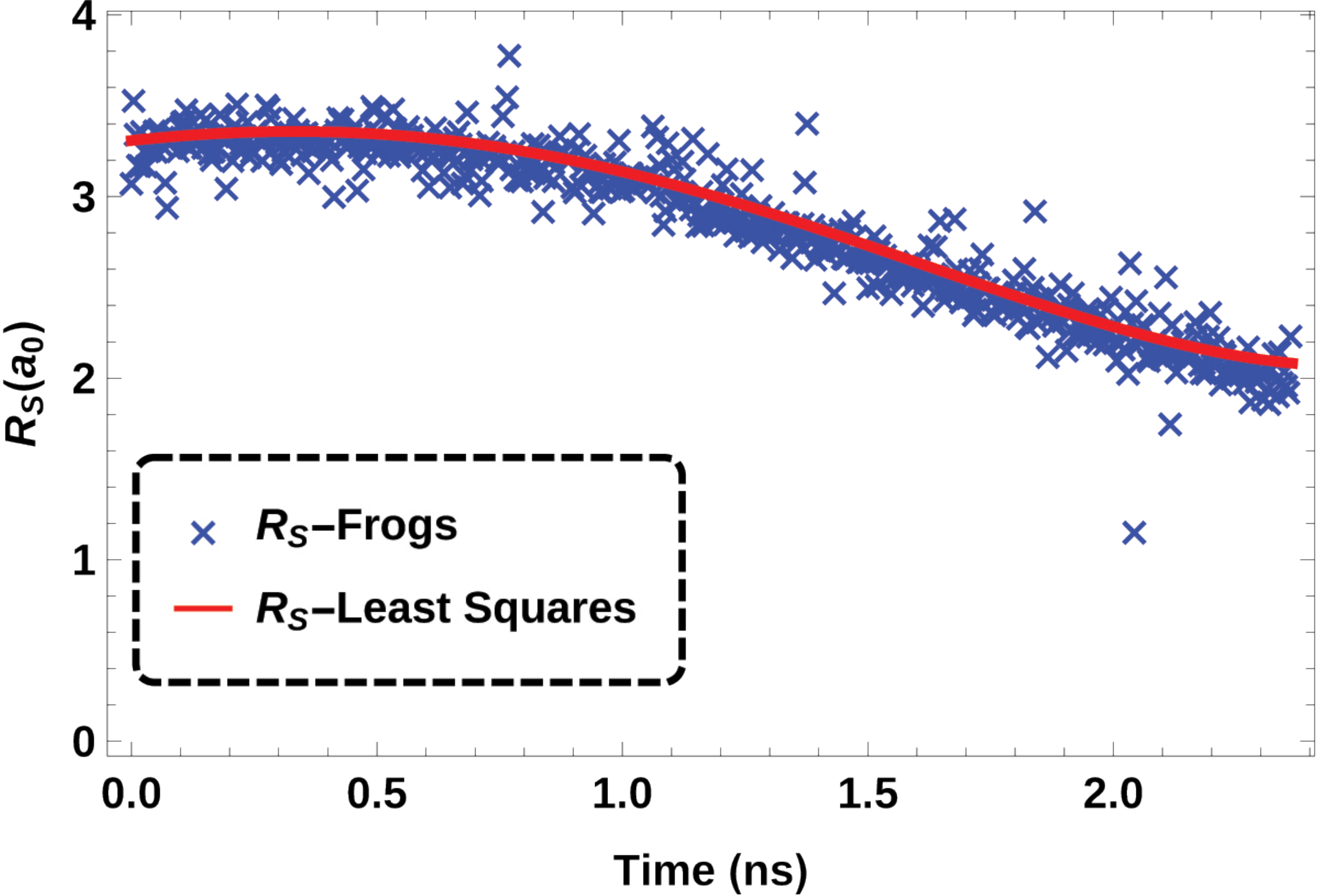}
    \caption{Evolution of $R_S$ for the initial 2.5 ns of dynamics simulation using model III. The graph shows that the skyrmion radius decreases in time, just as the skyrmion approaches the edge of the nanoribbon. This is a good example of dynamically changing behavior that can be thoroughly quantified using the Frogs Method. The cross markers show data obtained by the Frogs while the line shows a least squares fitting of this same data.}
    \label{fig:Rs_vs_t}
\end{figure}

During the total course of simulating with model I, the initial configuration evolved to $R_S=(3.25\pm0.11)a_0$ and $s=(0.38\pm0.04)$. For model II, these values are $R_S=(3.33\pm0.04)a_0$ and $s=(0.40\pm0.01)$, and for model III, $R_S=(3.01\pm0.36)a_0$ and $s=(0.36\pm0.05)$. The larger deviation in model III comes from a temporal dependence of $R_S$ through the dynamics, in which the skyrmion radius shrinks as it approaches a boundary. The same reduction  in the proximity of a boundary can be seen in $s$, but with smaller deviation. This behavior is shown in figure \ref{fig:Rs_vs_t}.

From the $R_s(t)$ profile, it can be seen that the skyrmion radius reduces as the time passes and it approaches the nanoribbon edge. This effect is due to a skyrmion-edge repulsion which also reduces the skyrmion velocity as it starts to travel tangentially to the closest boundary. It has been modeled in stationary cases and equilibrium values have been observed experimentally and also qualitatively in simulations  \cite{WangSize,ZhangRepul}. Note however, that a properly quantified measurement of the change in skyrmion size, obtained in the course of the dynamics simulations to a high degree of precision, is neither readily accessible nor trivial from first principles and the magnetization profiles alone. This description thus immediately showcases the usefulness of the Frogs Method.

\section{Conclusion}
\label{conclusion}

We proposed a bio-inspired data extracting method to handle typical $\textbf{m}_i(t_n)$ data obtained from most of the traditional micromagnetism packages and to obtain skyrmion trajectory and information regarding internal coordinates. The so called Frogs Method is inspired by the search of food in a hypothetical frog army, which is in essence very similar to the traditional PSO, but brings a random walk approach to the problem that is comparable to the widely used Monte Carlo techniques. 

Our methodology has been compared with previously reported ones in handling output from skyrmion dynamics of three model structures, to obtain the trajectory of a skyrmion towards equilibrium. The Frogs Method has been able, in addition to properly modelling the trajectory, to also track data from equilibrium $R_S$ and $s$ values, as well as showing their behavior over time. This data pointed to a shrinking of $R_S$ over time that is clearly visible in the dynamics, but so far remained elusive to quantify in a readily accessible manner.

In the trajectories, the Frogs Method has shown to satisfiably track skyrmion position in comparison to the other methods. The value obtained by $m_z$ systematically gives positions deviated towards the center of the studied structures (by failing to account for boundary demagnetization), while the trajectory obtained by $n$ does the same in the opposite direction, towards boundaries (by failing to account for a small but non-negligible boundary topological charge). The Frogs Method on the other hand, gives a trajectory that shows a random error that is estimated to be around $0.1a_0$, significantly smaller than the errors from the other methods even in the visual impression from figure \ref{fig:frame50}. This error can be further reduced if required, by averaging over a larger number of measurements with different random numbers or even by using different adjustable parameters that better model the desired system.

We present a method that has been shown very helpful in obtaining data from micromagnetic simulations. Though the direct application to the particular case of skyrmion dynamics was shown, the method's fitting function and variables can be easily redefined in order to extract data from more general systems, including systems both inside and outside of the field of micromagnetism.

\section*{Acknowledgements}

The author acknowledges support by the Conselho Nacional de Desenvolvimento Cient\'ifico e Tecnol\'ogico (CNPq), the Funda\c c\~ ao de Amparo \`a Pesquisa do Estado de Minas Gerais (FAPEMIG), the Financiadora de Estudos e Projetos (FINEP) and the Coordena\c c\~ ao de Aperfei\c coamento de Pessoal de N\'ivel Superior (CAPES) for financial support, as well as UFJF for the infrastructure used.

\appendix

\section{Pseudo-code}
\label{pseudocode}

In order to illustrate the implementation of our proposed meta heuristics, we proceed to show a pseudo-code in which we minimize \texttt{f()}, that could be any function, not only the one that was used for the case of a skyrmion. The value of this function is evaluated for every frog \texttt{i}, for which the associated position is given by \texttt{x(i)}. At each step, we print \texttt{g} and \texttt{best}, representing the value of the function and the position for the best frog.
        
    \begin{verbatim}
    BEGIN
    
    READ{Rmax, phi, x0, dx, nstep, nfrog}
    
    INITIALIZE best=+inf
    FOR i < nfrog DO
        INITIALIZE x0-dx < x(i) < x0+dx
        CALCULATE f(i)
        IF f(i) < best THEN
            best = f(i)
            g = x(i)
        END IF
        fprevious(i)=f(i)
    END FOR
    
    FOR t < nstep DO
        FOR i < nfrog DO
            v(i)=Rmax/(phi+1)*(Rand(-1,1)
                 +phi*Rand(0,1)*sign(gx(i)))
            x(i)=x(i)+v(i)
        END FOR 
        FOR i < nfrog DO
            CALCULATE f(i)
            IF f(i) < best THEN
                best = f(i)
                g = x(i)
            END IF
            IF f(i) > fprevious(i) THEN
                x(i) = x(i) - v(i)
            END IF
        END FOR
        PRINT{g,best}
        FOR i < nfrog DO
            fprevious(i)=f(i)
        END FOR
    END FOR
    
    END
    \end{verbatim}
        
In addition, this code can be adapted for maximization problems or for higher dimensions by changing some \texttt{IF} inequalities or making \texttt{x(i)} a matrix variable. The code can be even further generalized in more abstract forms of \texttt{x(i)}; the only condition is to have a suitable  fitting function \texttt{f(i)} to treat the problem. In the particular case of this work, $\texttt{f}:\mathbb{R}^n \to \mathbb{R}$, but the approach we demonstrated can be adapted to larger and even more complex optimization problems.

\bibliographystyle{model1-num-names}

\end{document}